\documentclass[aps,twocolumn,showpacs,floatfix]{revtex4}
\usepackage[T1]{fontenc}
\usepackage{graphicx}
\newcommand\forget[1]{}
\newcommand\comment[1]{}

\begin{document}


\title{Towards a monolithic optical cavity for atom detection and manipulation}


\author{S. Gleyzes}
\email{sebastien.gleyzes@institutoptique.fr}
\homepage{http://www.atomoptic.fr/}
\affiliation{Laboratoire Charles Fabry, Institut d'Optique, CNRS et Universit\'{e} Paris-sud, Campus Polytechnique, 91127 Palaiseau CEDEX, France}
\author{A. El Amili}
\affiliation{Laboratoire Charles Fabry, Institut d'Optique, CNRS et Universit\'{e} Paris-sud, Campus Polytechnique, 91127 Palaiseau CEDEX, France}
\author{R. A. Cornelussen}
\affiliation{Laboratoire Charles Fabry, Institut d'Optique, CNRS et Universit\'{e} Paris-sud, Campus Polytechnique, 91127 Palaiseau CEDEX, France}
\author{P. Lalanne}
\affiliation{Laboratoire Charles Fabry, Institut d'Optique, CNRS et Universit\'{e} Paris-sud, Campus Polytechnique, 91127 Palaiseau CEDEX, France}
\author{C. I. Westbrook}
\affiliation{Laboratoire Charles Fabry, Institut d'Optique, CNRS et Universit\'{e} Paris-sud, Campus Polytechnique, 91127 Palaiseau CEDEX, France}
\author{A. Aspect}
\affiliation{Laboratoire Charles Fabry, Institut d'Optique, CNRS et Universit\'{e} Paris-sud, Campus Polytechnique, 91127 Palaiseau CEDEX, France}


\author{J. Estève}
\altaffiliation{Present address: LKB, 24, rue Lhomond, 75231 Paris CEDEX 05, France}
\affiliation{Laboratoire de Photonique et de Nanostructures, CNRS, 91460 Marcoussis, France}

\author{G. Moreau}
\altaffiliation{Present address: EDF R\&D, 6 quai Watier, 78401 Chatou, France}
\affiliation{Laboratoire de Photonique et de Nanostructures, CNRS, 91460 Marcoussis, France}

\author{A. Martinez}
\affiliation{Laboratoire de Photonique et de Nanostructures, CNRS, 91460 Marcoussis, France}

\author{X. Lafosse}
\affiliation{Laboratoire de Photonique et de Nanostructures, CNRS, 91460 Marcoussis, France}

\author{L. Ferlazzo}
\affiliation{Laboratoire de Photonique et de Nanostructures, CNRS, 91460 Marcoussis, France}

\author{J.C. Harmand}
\affiliation{Laboratoire de Photonique et de Nanostructures, CNRS, 91460 Marcoussis, France}

\author{D. Mailly}
\affiliation{Laboratoire de Photonique et de Nanostructures, CNRS, 91460 Marcoussis, France}

\author{A. Ramdane}
\affiliation{Laboratoire de Photonique et de Nanostructures, CNRS, 91460 Marcoussis, France}


\date{\today}
\begin{abstract}
We study a Fabry-Perot cavity formed from a ridge waveguide on a AlGaAs substrate.
We experimentally determined the propagation losses in the waveguide at 780~nm, the wavelength of Rb atoms.
We have also made a numerical and analytical estimate of the losses induced by the presence of the gap which would allow the interaction of cold atoms with the cavity field.
We found that the intrinsic finesse of the gapped cavity can be on the order of $F\sim 30$,
which, when one takes into account the losses due to mirror transmission,
corresponds to a cooperativity parameter for our system $C\sim 1$.\end{abstract}


\maketitle

Recent years have seen the successful detection of single atoms on an atom chip using the interaction of  an atom with a high finesse optical cavity \cite{Teper06,Trupke07,Colombe07}. These experiments represent great strides forward in our ability to perform cavity quantum electrodynamics experiments in the versatile and compact environment of an atom chip. For the purposes of atom detection, a important figure of merit is the cooperativity parameter, $C = g^2/\kappa\gamma$, where $\gamma$ is the natural atomic half-linewidth, $ \kappa$ is the cavity half-linewidth and $g$ is the ``atom coupling to the cavity'', that is the Rabi frequency corresponding to a single photon in the cavity. Roughly speaking, the quantity $C^{-1}$ corresponds to the mean number of spontaneously scattered photons necessary to detect a single atom with a signal to noise of unity. The quantity $C$ is proportional to the square of the transition wavelength and to the finesse of the cavity, and inversely proportional to the cross sectional area of the cavity mode. The miniaturization of the optics for this type of experiment is important, in part because the small mode area can lead to significant cooperativity, even without a a large finesse \cite{Haase06}. The use of a low finesse, much less sensitive to external disturbances, leads to a simplification of the experimental setup needed to stabilize the cavity.

Other workers in this field have already demonstrated small cavities with impressive finesse
\cite{Teper06,Trupke07, Colombe07, Trupke05, Steinmetz06, Wilzbach06}.
However, these cavities often employ optical elements glued onto a traditional atom chip.
In this paper we discuss the possibility of building a microcavity based on an optical waveguide integrated on a GaAs substrate. Integrated optics has already been successfully used to implement photonic quantum gates \cite{Politi08}. Combined with atom chip technology, it would allow us to directly fabricate on the atom chip substrate both the microwires necessary to trap and manipulate cold atoms and the optical cavity used to detect them. The atomic and optical waveguide would be aligned using lithography techniques to the nanoscale, which would enable the parallel construction of hundreds or thousands of such cavities on a single chip.
This feature promises to be of considerable interest if quantum computation
using neutral atoms is ever to become a reality \cite{Trupkeproposal07}.
The choice of a semiconductor waveguide would allow one to tune the cavity resonance without any mechanical adjustment, by changing the charge carrier density \cite{Huang90}, making the setup very robust. In addition, with the use of GaAs one can envision the integration on the substrate of active elements such as the photodetectors and the laser sources involved in the experiment.

\begin{figure}
 \includegraphics[width=\linewidth]{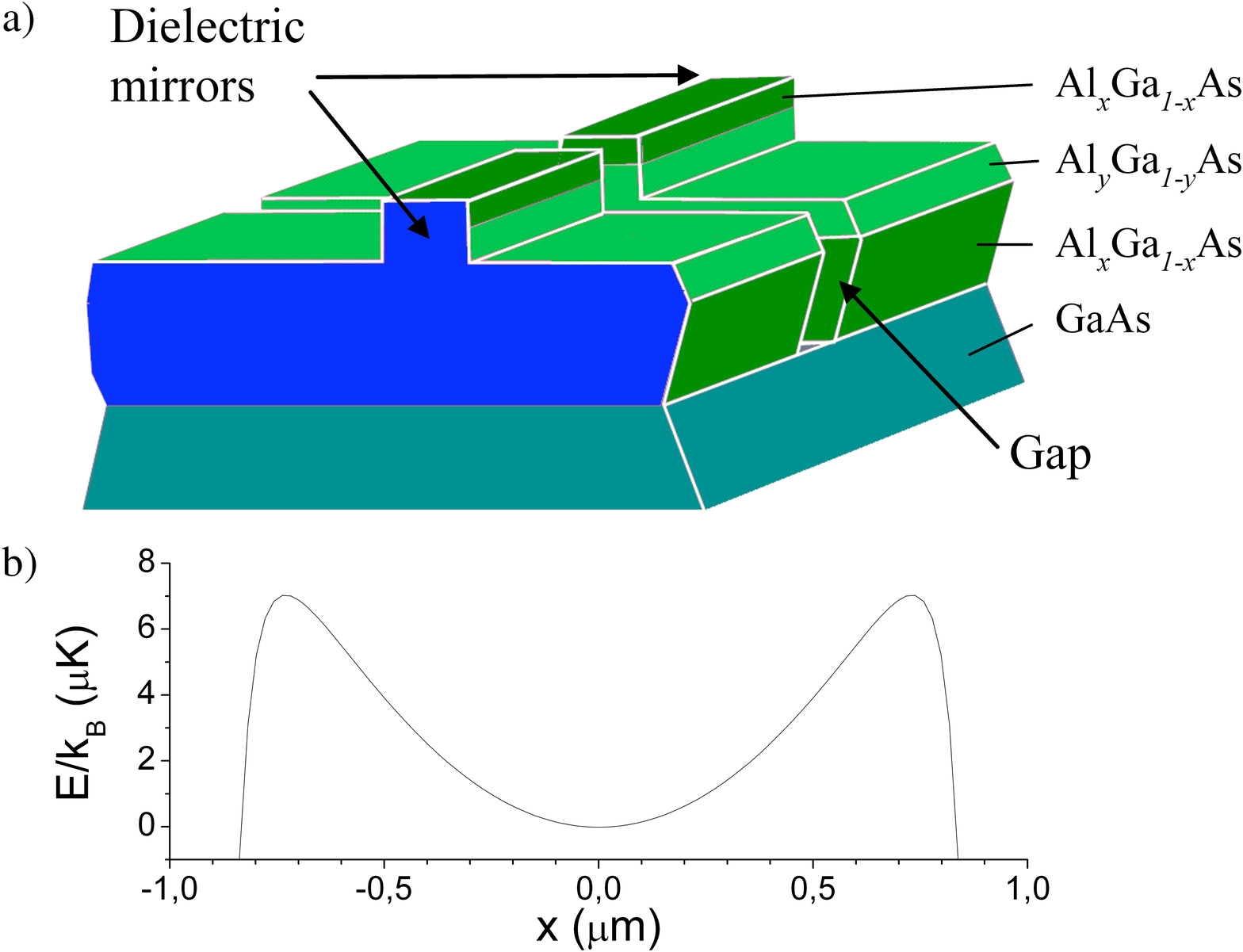}\\
  \caption{a. A schematic diagram of the cavity. b. Typical trapping potential inside the gap, taking into account the magnetic trap (the trapping frequency for $^{87}$Rb is $\omega_{\mathrm{trap}}/2\pi= 9$~kHz) and the Casimir-Polder potential for gap of width $d=2$~$\mu$m. }\label{fig-schematic}
\end{figure}

A schematic diagram of such a cavity is shown in Fig.~\ref{fig-schematic}.a.
It is formed by a waveguide both ends of which are coated with a highly reflecting coating.
In the middle of the waveguide, a gap allows the atoms to interact with the field of the cavity.
As already illustrated in Ref.\cite{Lin04}, long-range van der Waals-type interactions between the atom and the surface impose a minimum width on this gap, otherwise no atom trapping potential can exist in the gap.
In our case, we plan to confine atoms magnetically on an atom chip.
The typical strength of such a trap imposes a width of 2~$\mu$m (see Fig.~\ref{fig-schematic}.b).
We will first describe the geometry of our optical waveguide, and our measurement of the propagation losses. We will then discuss the losses induced by the gap in the cavity, and finally give the estimate of the cooperativity that one could hope to attain with this system.

\begin{figure}
 \includegraphics[width=\linewidth]{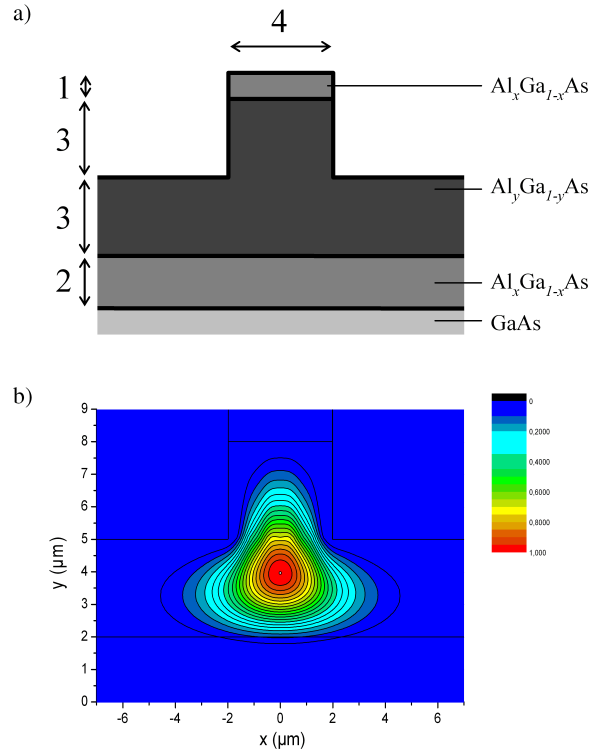}\\
  \caption{a. Cross-section of the AlGaAs waveguide showing the critical dimensions (in microns). b. Numerical simulation of the mode intensity profile at 780 nm.}\label{fig-guide}
\end{figure}

To limit the absorption of the optical waveguide at the resonant wavelength of Rb, 780 nm, we use a AlGaAs structure with a high concentration of Al \cite{Pavesi94,Ariza-Calderon98}.
The rather large gap in the guide in turn imposes the size of the waveguide:
in order to avoid diffraction losses when the light propagates in the gap,
the waveguide itself must be large.
Reference \cite{Horak03} for example reached the same conclusion for a cylindrical waveguide.
A cross section of the guide is shown Fig. \ref{fig-guide}.a.
We first deposit three AlGaAs layers:
Al$_x$Ga$_{1-x}$As/Al$_y$Ga$_{1-y}$As/Al$_x$Ga$_{1-x}$As
on a GaAs substrate using molecular beam epitaxy.
The Al concentrations, $x=79.5\%$ and $y=77.5\%$,
are chosen so that the refractive index at $\lambda=780$ nm
of the middle layer, $n_y=3.155$, is slightly higher than that of the top and bottom layers, $n_x=3.145$. The light is thus confined in the plane by total internal reflection.
The two top layers are then etched away using SiCl$_4$/O$_2$
reactive ion etching (RIE) to create the 4 $\mu$m $\times$ 4 $\mu$m ridge. Using numerical simulation software (ALCOR) we can calculate the profile of the eigenmode of the waveguide (Fig. \ref{fig-guide}.b), and the mode area $A=9.9$~$\mu$m$^2$.

To characterize the propagation losses in the waveguide, we coat the ends of the waveguides with dielectric mirrors, and measure the finesse of the resulting cavity. Each mirror is a stack of 3 to 6 pairs of YF$_3$/ZnS layers each with quarter wavelength optical thicknesses.
The theoretical reflectivities of the mirrors are: $R_3= 91.3\%$ and $R_6= 99.4\%$.

\begin{figure}
  \includegraphics[width=\linewidth]{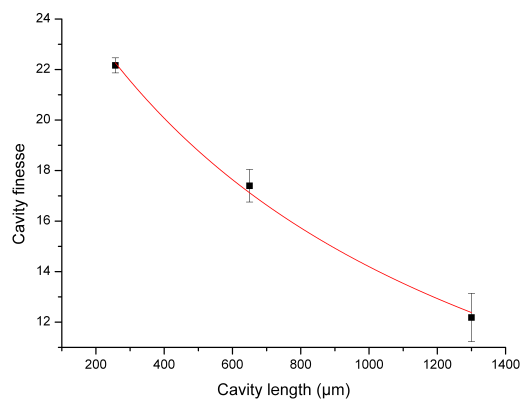}\\
  \caption{Best measured finesse as a function of cavity length for a 3-pair coating.
  We use the data to determine the attenuation coefficient in the guide. (See text.)}\label{fig-finesse1}
\end{figure}

The propagation loss measurement was performed in two steps: we first coated three samples cleaved to three different lengths ($l_1=260$ $\mu$m, $l_2=650$ $\mu$m and $l_3=1300$ $\mu$m) with a 3 pair mirror on each side, and measured the finesse of each cavity.
For each length $l_i$, the finesse of the cavity is given by
\begin{equation}
\mathfrak{F}_i = \frac{\pi\sqrt{R_3 e^{-\alpha l_i}}}{1- R_3 e^{-\alpha l_i}}
\end{equation}
where $\alpha$ characterizes the propagation losses of the waveguide.
A fit to the data allows us to estimate $R_3^{exp} = 89 \%$ (in good agreement with the theoretical value) and $\alpha = 1.07\pm0.06$ cm$^{-1}$ (Fig. \ref{fig-finesse1}).

\begin{figure}
  \includegraphics[width=\linewidth]{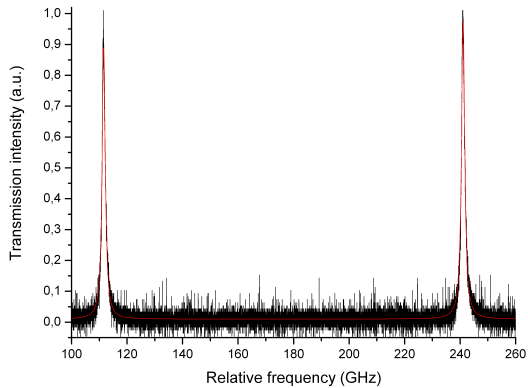}\\
  \caption{Transmission of the 330 $\mu$m cavity coated with 6 pair mirrors ($R_6= 99.4\%$). The width of the resonance is $2\kappa = 2\pi\cdot1.4$~GHz, with a free spectral range of $129$~GHz, corresponding to a finesse of $\mathfrak{F}_{intr}=92$.}\label{fig-finesse2}
\end{figure}

To confirm this measurement, we deposited six YF$_3$/ZnS pairs on each face of a fourth sample with length $l_4=330$ $\mu$m.
 The theoretical value of $R_6$ is much closer to unity than the propagation loss on a single trip $e^{-\alpha l_4} \approx 96 \%$, so that this time the cavity finesse is limited by the waveguide absorption.
We measured a linewidth of the resonance $2\kappa = 2\pi\cdot 1.4\pm0.1$ GHz.
Knowing the effective refractive index of the waveguide $n=3.50\pm0.04$, we can make another determination of the attenuation, $\alpha = 1.03\pm0.06$ cm$^{-1}$.
For a 330 $\mu$m cavity, this corresponds to an intrinsic finesse $\mathfrak{F}_{intr}=92$ (Fig. \ref{fig-finesse2}).
The source of the propagation loss could be either scattering by the surface roughness of the
waveguide or by inhomogeneities in the bulk, or an absorption process within the material.
Our data do not distinguish these possibilities.

Of course, in such a cavity the electric field is confined inside the semi-conductor material.
If we want to let the atom interact with the field, we need to open a gap in the waveguide.
The constraint that the gap be at least $2~ \mu$m in width means that the gap
contributes additional loss to the cavity since the light is not confined and will diffract
in propagating across the gap.
This type of effect was discussed in the Ref. \cite{Horak03} for a
cylindrical waveguide cross section with a gaussian beam profile.
The authors found a single pass loss of order 0.12 \% for a $2~\mu$m gap. In our case, the losses are more severe for two reasons. First, the eigenmode of the waveguide is non gaussian and diffracts faster during the free-space propagation than in \cite{Horak03}. Furthermore, its refractive index is much larger than the refractive index of a fiber. The Fresnel reflections at the gap interface are higher, which enhances the field amplitude in the gap and therefore increases the losses. We have estimated the diffraction effect numerically as described in the appendix, and found losses of order 7\% for a single pass. This limits the cavity finesse to a value of 40,  neglecting any other loss in the waveguide.
If we include the propagation loss in a $300~\mu$m cavity the finesse is limited to 30.

For the purpose of maximizing the interaction between the atom and the field in such a cavity, the real figure of merit of the cavity is not the finesse, but rather the cooperativity discussed in the introduction.
To estimate this quantity, it is necessary to take into account the high index of refraction of the guide which renders the reflection at the interfaces of the gap of order 27\%.
The air gap therefore
produces a low finesse Fabry-Perot cavity of its own.
If the length of the gap is adjusted so that it corresponds to an integer number of half-wavelengths,
the Fabry-Perot transmission is equal to unity (neglecting the losses inside the gap).
However, the field inside the gap is enhanced by constructive interference of the multiple reflections at the gap interfaces.
If the total length of the cavity is properly chosen, the amplitude of the electric field inside the gap (and therefore seen by the atom) is a factor of $n$ larger than the electric field in the rest of the cavity \cite{Horak03}, which improves the cooperativity by a factor $n^2$ ($\sim 10$ in our case) with respect to a free-space Fabry Perot cavity of same finesse and mode area.

Finally, one can estimate the cooperatity of such an integrated cavity. Knowing the mode area and the length of the cavity ($L=300$~$\mu$m), we can calculate the coupling factor $g/2\pi = 120$~MHz. The intrinsic finesse of the cavity with gap is 30, which corresponds to an intrinsic loss rate $2\kappa_{intr}=2\pi\cdot 4.8$~GHz.
The total cavity loss $\kappa $ is the sum of the
intrinsic loss and the loss due to the transmission of the mirror $\kappa_{T}$.
To optimize the single-atom detection signal to noise, one has to choose  $\kappa_{T}=\kappa_{intr}$, leading to a total cavity loss $\kappa/2\pi = 4.8$~GHz.
Using the half width for the $D2$ line of rubidium $\gamma/ 2\pi = 3$~MHz,
we can expect a cooperativity $C=g^2/\kappa\gamma \sim 1$.

We have described a new type of micro-cavity that can be integrated onto an atom chip substrate for direct detection of atoms trapped near the surface.
The cavity is based on an integrated waveguide grown on a GaAs substrate.
The propagation losses at 780 nm are small enough to reach a finesse on the order of 100 for a 300~$\mu$m cavity, and could be further reduced by improving the fabrication process or optimizing the Al concentration in the semiconductor.
To let the atom interact with the cavity field, we plan to open a gap in the middle of the cavity.
The realization of the gap remains a technical challenge, but
our theoretical estimate of the losses induced by the presence of the gap
shows that they remain low enough to reach a cooperativity on the order of unity.
Thus one can envision the
construction of completely integrated devices where hundreds of atom waveguides and cavities are nano-fabricated in parallel on the same substrate.

 The atom optics group is member of l'Institut Francilien de la Recherche sur les Atomes Froids. This work has been supported by the EU under grants MRTN-CT-2003-505032, IP-CT-015714, and by the CNANO  program of the Ile de France.

 \appendix

 \section{Numerical estimate of the losses in the gap}

The losses introduced by the presence of the gap of width $d$ are due to the fact that the mode inside the gap is no longer confined and will diffract during its free space propagation from one interface to the other, and some of the energy cannot be recoupled back into the second half of the cavity. Since we cannot calculate the free-space propagation of the mode analytically, we had to perform numerical simulations.

To estimate the diffraction losses, a first approximation is to compute the overlap $Q$ between the mode after a free-space propagation over a distance $d$ and the eigenmode of the waveguide. The cavity loss rate is then the energy flow inside the gap (calculated as if the coupled-cavity system were lossless) multiplied by $(1-|Q|^2)$, the fraction of the energy that is not coupled back into the waveguide after a single trip across the gap \cite{Horak03}. However, this approximation holds only as long as the amplitude of the light reflected at the air/guide interface is small. Otherwise, while the light bounces back and forth across the gap, its transverse profile keeps expanding, and the overlap with the eigenmode of the waveguide gets smaller. The fraction of the energy not coupled back into the cavity thus becomes larger than $(1-|Q|^2)$.

In our case, because of the high refractive index of the waveguide material, the Fresnel reflection at each interface is not negligible. Therefore, to estimate the diffraction losses, we have to simulate numerically the losses induced by the gap taking into account the multiple reflections. First we consider the transmission and reflection of a simple gap of length $d$. The reflection and transmission coefficients have been calculated with a 3D fully-vectorial aperiodic-Fourier modal method \cite{Hugonin05}. We calculate the eigenmode of the optical waveguide, and then compute the propagation of this mode in a wave\-guide - air - wave\-guide stack. We denote by $R$ and $T$ the intensity reflection and transmission coefficients.
We observe that they exhibit oscillations as the length $d$ increases, corresponding to the Fabry Perot interferences between the two reflections on each side of the gap. The fraction of the energy that is lost in the gap is given by $1 - R - T$.
This fraction increases as the length of the gap increases, exihibiting periodic maxima
when the length of the gap corresponds to a integer number of half-wavelengths.
Qualitatively, this behavior is easily understood, since the diffraction loss is proportional to the intensity of the field in the gap, and is therefore maximal when the interferences in the gap are constructive, that is when the small cavity formed by the gap is resonant with the optical field \cite{Wilzbach06}.
Our goal however, is to couple the atoms to the electromagnetic field of the optical waveguide,
and therefore we choose to maximize the field inside the gap in spite of the increased loss.

\begin{figure}
   \includegraphics[width=1.\linewidth]{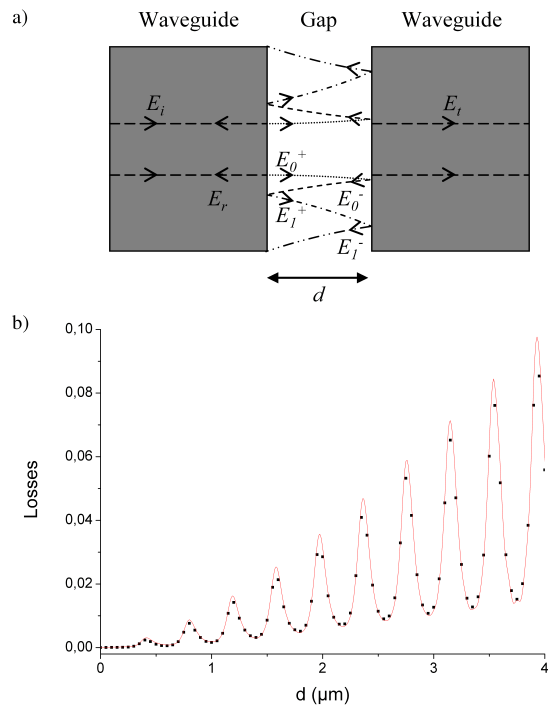}\\
  \caption{a. Diagram of the waveguide - gap - waveguide stack used for the numerical simulation and analytic model. The $E^\pm_p$ fields correspond to the successive reflections of the incident field $E_i$ through the gap. Since the mode is not confined in the gap, it diffracts further and further while bouncing back and forth on the gap interfaces.
   b. Energy loss in the gap, defined as $1 - (R + T)$, as a function of the gap width $d$.
The squares show the results of the numerical simulation and the solid line is the analytical model.}\label{fig-app-fit}
\end{figure}

To understand the behavior of the losses in the cavity a little more quantitatively,
we developed a simple semi-analytical model for the field in the gap.
We consider the mode inside the cavity as the superposition of $E^+_p$ and $E^-_p$,
the fields corresponding to the $2p^{\mathrm{th}}$ (propagating from left to right) and $2p+1^{\mathrm{th}}$ (propagating from right to left) reflections respectively at  the gap interfaces
(see Fig \ref{fig-app-fit}.a).
The amplitude of $E^+_p$ ($E^-_p$) can be expressed as a function of the incident field amplitude $E_i$ as $tr^{2p}E_i$ ( $tr^{2p+1}E_i$), where $r$ and $t$ are the reflection and transmission coefficient of the light amplitude at a single waveguide/air interface.
In first approximation, we use the bulk Fresnel coefficients for $r$ and $t$.

The field $E_t$  transmitted through the gap is therefore the sum of the transmission of all the $E^+_p$ components of the field \begin{equation}
E_t = t \sum_p Q^+_p t r^{2p} E_i .
\end{equation}
The factor $Q^+_p$ takes into account the fact that, due to diffraction, the mode profile of the field after the multiple reflections inside the gap does not exactly match the eigenmode of the waveguide, and some of the light cannot be coupled back into the right part cavity. To estimate $Q^+_p$, we compute the
free-space propagation of the waveguide eigenmode $E_0$, and assume that the transverse profile of  $E^+_p$ at the right gap interface is the same as the transverse profile of the waveguide mode $E_0$ after propagating a distance $(2p+1)d$ in free space.

In a similar way, one can write the field reflected by the gap as the
sum of the light directly reflected by the waveguide/air interface $ -r E_i$
and the mode-matched part of $E^-_p$ that is transmitted back into the left part of the waveguide.
\begin{equation}
E_r = -r E_i + t \sum_p Q^-_p t r^{2p+1} E_i .
\end{equation}
Again, we estimate the factor $Q^-_p$ by computing the overlap between the
eigenmode of the waveguide $E_0$ and the profile of $E_0$ after a free-space propagation
of $2(p+1)d$.

The theoretical reflection and transmission coefficients for the intensity are given by $R=|E_r/E_i|^2$ and $T=|E_t/E_i|^2$. We can then compare the loss $1 - R - T$ with the numerical simulations, and find an excellent agreement (Fig. \ref{fig-app-fit}.b).

Finally, to estimate the contribution to the finesse due to the losses in the gap, we need to calculate the amplitude attenuation factor $r_{rt}$ of the light during a round trip inside the cavity.
Therefore we numerically compute the amplitude of the light reflected back by the
ensemble consisting of the optical waveguide on the left in Fig.~\ref{fig-app-fit}.a,
a gap of length $d=1.96$~$\mu$m (5 half wavelengths at 780 nm),
and a second optical waveguide (on the right) terminated with a 100\% reflector.
Depending on the length $L$ between the gap and the perfect mirror, the multiple reflections inside the gap of the light coming from the left waveguide interfere constructively or destructively with the multiple reflections of the light coming back from the right waveguide.
The loss induced by the gap is maximal when the interference is constructive ($r_{rt}=0.93$), but the amplitude of the electromagnetic field in the gap is enhanced by a factor $n$ \cite{Horak03}.
If the interference is destructive, the loss is reduced ($r_{rt}>0.99$) but the coupling to the atoms in the gap is also smaller. In this paper we consider the case of constructive interference to gain the $n^2$ factor on the cooperativity, but a more detailed analysis of how this might be improved by changing various parameters will be published elsewhere.

\end{document}